Submission type: Article

# Epitaxial phases of high Bi content GaSbBi alloys

*Joonas Hilska[1,*], Eero Koivusalo[1], Janne Puustinen[1], Soile Suomalainen[1], and Mircea Guina[1]*

[1]Optoelectronics Research Centre, Tampere University of Technology, Korkeakoulunkatu 3, 33720, Tampere, Finland

**E-mail**: joonas.hilska@tut.fi ([*]corresponding author)

**Abstract**: GaSbBi alloys have recently emerged as attractive materials for mid-infrared optoelectronics owing to strong band gap reduction enabled by Bi incorporation into the GaSb matrix. The fundamental understanding of the epitaxial process required to demonstrate high quality crystals is in an early-developmental phase. From this perspective, we report on the key role played by the Sb/Ga flux ratio in controlling the structural quality and incorporation of high Bi content (up to 14.5 %-Bi), revealing three distinct epitaxial phases. The first phase (below stoichiometric Sb/Ga) exhibits Ga-Bi compound droplets, low crystal quality, and reduced Bi content. At the second phase (above stoichiometric Sb/Ga), the crystal exhibits smooth surfaces and excellent crystallinity with efficient Bi incorporation. The last phase corresponds to exceeding a Sb/Ga threshold that leads to reduced Bi incorporation, Bi droplets and degraded crystallinity. This threshold value that defines the optimal growth window is controlled by the temperature as well as the Bi/Ga ratio. Increasing temperature increases the threshold, albeit simultaneously reducing Bi incorporation. Conversely, increasing the Bi/Ga flux ratio increases Bi incorporation, while narrowing down and ultimately closing the window. This study provides a general framework enabling development of high quality GaSbBi heterostructures for emerging mid-infrared optoelectronics.

# 1. Introduction

The mid-IR 2–5 µm spectral region is becoming increasingly important for photonics owing to a wealth of emerging application areas, ranging from medical diagnostics [1–3] to spectroscopy [4, 5] and even communications [6, 7]. At this wavelength region, availability of strong fundamental rotational and vibrational molecular transitions allows selective spectroscopic detection of a large number of molecules. For example, of particular importance are transitions at around 2.33–2.78 µm associated with HF, 2.5–2.75 µm for $H_2S$, ~2.65 µm and 4.2–4.3 µm for $CO_2$, or 3.2–3.45 µm for $CH_4$ [8], which enable real-time monitoring of environmental pollutants and industrial processes. The advance of mid-IR photonic applications is rapidly gaining momentum owing to emergence of photonic integration solutions enabling deployment of a complex system in a very small foot-print using cost-effective and scalable technology platforms [9, 10]. However, these advances in integration and application aspects has to be paralleled by developing suitable mid-IR optoelectronic materials for light generation, an area that has been lacking much behind the more mature counterparts at visible and near-infrared wavelengths.

In general, the challenges in band-gap engineering of mid-IR friendly semiconductor materials has led to the development of complex approaches for light sources, such as use of quaternary/quinary compounds or cascade heterostructures, which still face several physical limitations [11, 12]. For example, the cascade laser design requires intricate multistage superlattices, while still facing issues such as high power consumption in quantum cascade lasers and poor temperature behavior in interband cascade lasers. These features are indeed not compatible with requirements for cost-effective photonic integration platforms to power large-scale applications deployment. On the other hand, extending the operation wavelength of conventional type-I GaInAsSb quantum well (QW) lasers over ~3.4 µm is severely impaired by insufficient valence band offsets leading to poor hole confinement [13–15]. From this perspective, newcomers in the mid-IR alloy family such as GaSbBi are of high interest. Notably, GaSbBi exhibits advantageous properties, such as large band gap bowing and enhanced spin-orbit splitting [16–19], which promise to mitigate issues in current GaSb-based technologies. In particular for GaSbBi, the valence band offset increases with Bi content owing to the valence band anti-crossing effect [18]. Besides this, GaSbBi has unique advantages in inhibiting specific Auger recombination and intervalence absorption paths [20], while also enabling use of simple low-loss Al-free waveguides with improved mode confinement. Despite these well-recognized advantages, mid-IR GaSbBi heterostructures are at an early stage of development, with the first demonstration of a GaSbBi-based laser diode emitting at 2.7 µm using 11.5 %-Bi QWs made only recently [21]. Therefore, an open question that has to be addressed relates to defining the epitaxial regime that leads to high quality GaSbBi alloys with sufficiently high Bi content (e.g. >14 %-Bi [19]) enabling emission in the 3–5 µm range.

Generally speaking, GaSbBi alloys belong to the metastable III-V-Bi family, where unconventional epitaxial conditions, i.e. low growth temperatures and near stoichiometric V/III flux ratios, are required to incorporate Bi into the host material [22]. For the most extensively studied III-V-Bi alloy, GaAsBi [23–25], growth at these conditions has revealed temperature and Bi related defect incorporation to hinder the optical quality [26] and make fabrication device-grade high Bi content (e.g. >10 %-Bi) alloys challenging due to need for careful As/Ga flux control [25, 27, 28]. Considering the GaSb-based system, the effects of low temperature and high Bi incorporation are not necessarily as detrimental; evidenced by room-temperature photoluminescence exhibited by (i) GaSb/GaInAsSb QWs grown at temperatures as low as 350 °C [11] and (ii) up to 14 %-Bi containing (albeit structurally impaired) GaSbBi alloys [19]. Regarding the As/Ga flux sensitivity of GaAsBi alloys, a recent study exploiting combinatorial growth [29] has pointed to distinct epitaxial regimes controlled by the As/Ga ratio, such that optimization of the material properties requires extreme As-flux control. However for GaSbBi, we note that the vapor pressure of Sb is much lower than for As, making it more controllable in growth and thus alleviating issues in Sb/Ga sensitivity and reproducibility. Despite these benefits, only few groups have dedicated efforts to explore the growth parameter space of GaSbBi and thus the intricate relationships between epitaxial parameters and resulting material properties are poorly understood or even unexplored.

In this paper, we apply a holistic approach to investigating the relationships between epitaxial conditions and material properties of high Bi-content (up to 14.5 %-Bi) GaSbBi alloys using combinatorial molecular beam epitaxy (MBE) [30]. The idea for combinatorial growth is to generate a large variety of growth conditions at different sample locations within a single growth run. Here, this is achieved by stationary (non-rotating substrate) growth that leads to a range of flux conditions created over the substrate, due to the inherently non-uniform molecular beams. The flux distributions are experimentally determined using separate calibration samples. These distributions are then correlated with the properties of the GaSbBi films, allowing us to map the growth parameter space with a high number of experimental points. The optimal growth windows are identified for different Bi incorporation ratios in terms of crystal and surface quality over a wide range of parameters. Finally, we explore the fundamental limits of Bi concentration while maintaining structural integrity within our parameter space and shortly discuss countermeasures for extending this limit.

## 2. Methods

The samples were grown in a conventional ten-port V80H MBE-reactor fitted with standard effusion sources for Al, Ga and Bi, while $As_2$ and $Sb_2$ were provided by valved cracker cells. The flux distributions were experimentally determined by stationary growth and ex-situ analysis of calibration structures. These structures were chosen as in our previous work [29], i.e. (i) an AlAs\GaAs heterostructure, (ii) Bi droplet epitaxy on GaAs(001), and (iii) an Sb-cap deposited at growth temperature $T_g$<40 °C for the Ga, Bi and Sb fluxes, respectively. Growth temperatures ($T_g$) herein refer to values given by a thermocouple situated behind the wafer.

The GaSbBi samples were grown on n-GaSb(001) 2" quarter-wafers. Native oxide removal was performed at $T_g$=700 °C for 10 min followed by a 150 nm GaSb buffer at 580 °C. The samples were then ramped down to the target growth temperature after which the sample rotation was stopped and set to a chosen constant position. The Sb-valve was closed midway through the ramp at $T_g$=455 °C to prevent Sb condensation on the growth surface. Three sets of samples were grown with increasing Bi/Ga beam equivalent pressure ratios (BEPR): 0.08, 0.11 and 0.15 (sets A, B and C). Each set comprised five samples grown at different $T_g$ ranging from 300 to 400 °C with a 25 °C step. We estimate the real surface temperature range to be 280 to 360 °C based on extrapolated pyrometer data. The nominal thickness and growth rate for the GaSbBi films were 150 nm and 0.5 µm h$^{-1}$ at the center of the wafer.

The samples were analyzed with high-resolution x-ray diffraction (HR-XRD), scanning electron microscopy (SEM), and atomic force microscopy (AFM). Thickness and composition for layers with good crystal quality was determined by fitting dynamical simulations to the HR-XRD measurements. The layers were assumed fully strained and a lattice constant of 6.27 Å was used for the GaBi binary [16]. For poor crystal quality layers, the composition was estimated from the angular mismatch of the GaSbBi and GaSb (004)-peaks. The surfaces were characterized by SEM in secondary electron (SE) and backscattered electron (BSE) modes with a Zeiss Ultra-55 instrument. AFM was measured in tapping mode with a Veeco Dimension 3100 microscope.

## 3. Results and discussion

Figure 1(a) shows a simplified schematic of the growth geometry where the flux distributions (figure 1(b)) create three surface phases (I, II, III) for a typical sample in this work. In phase I, the Sb/Ga flux ratio is below stoichiometric (Sb/Ga<1) which results into accumulation of submicrometer sized metallic droplets. SEM images from such droplets is shown in figure 1(c) for the sample in set B grown at 350 °C (i.e. moderate Bi flux and $T_g$). The SEM BSE image with Z-contrast suggests the droplets to consist of two parts: one part Ga (dark contrast) and one part Bi (bright contrast). Indeed, such compound droplets have been reported for GaSbBi [22, 31] as well as GaAsBi [28, 32, 33], and here are ascribed to the insufficient Sb/Ga ratio.

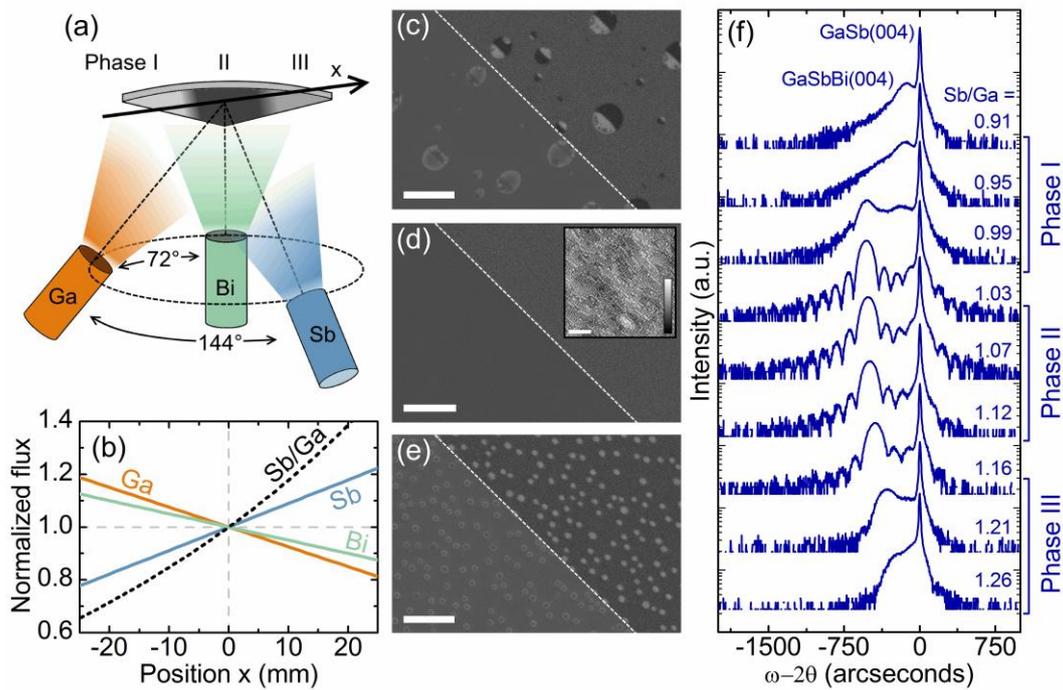

**Figure 1.** (a) Schematic of the flux distributions and resulting surface phases. (b) Experimentally determined flux distributions along the maximum Sb/Ga ratio gradient (indicated by the arrow in figure 1(a)). (c-e) SEM images from the sample grown at 350 °C in set B, where the bottom left and top right parts are imaged with SE and BSE detectors, respectively. The horizontal white scales correspond to 1 um, while the vertical scale in the AFM inset of (d) is 2 nm. (f) HR-XRD measurements over the same sample with estimated Sb/Ga values. The ω-2θ axis is centered to the GaSb(004) reflection.

As the Sb/Ga flux ratio is increased over the wafer, the size and density of the droplets decrease, to a point that a droplet-free surface is formed, i.e. phase II (Sb/Ga≥1). Figure 1(d) shows both SEM and AFM images from this phase, which show a smooth (RMS roughness of 0.35 nm) homogeneous surface with nm-scale undulations. Such undulations are typical for GaSbBi [16, 34] and are thought to be induced by presence of Ehrlich-Schwoebel diffusion barriers and anisotropic surface diffusion processes due to low temperature [35, 36].

Increasing the Sb/Ga flux ratio further, again leads to droplet formation, i.e. phase III. Figure 1(e) shows SEM images from this phase, where the bright BSE contrast of the droplets suggest they consist of Bi. These droplets are below 100 nm in radius and share similar facetted shape as in the Bi calibration sample, which has also been observed for Bi droplets on GaAsBi [32]. In fact, Bi droplet formation has been widely reported for III-V-Bi alloys [32, 37, 38], including GaSbBi [16, 31], which here is ascribed to the excessive Sb/Ga ratio.

The Sb/Ga flux ratio also influences the Bi content and crystalline quality. Figure 1(f) shows a set of HR-XRD measurements from the same sample as in the surface phase analysis. At areas corresponding to phase I (Sb/Ga<1), the GaSbBi(004) peak intensities are generally lower and are missing Pendellösung thickness fringes, indicating degraded crystalline quality, inhomogeneous composition and/or poor interface quality. This coincidence of degraded HR-XRD features and Ga containing droplets in phase I is in agreement with a recent study, where Ga droplets were found to promote inhomogeneous Bi incorporation in GaAsBi [39]. Additionally, we find that the Bi content is significantly lower (~2 %-Bi) than in phase II, due to smaller angular separation in the epilayer peak, in correlation with earlier analysis of figure 1(c) containing unincorporated surface Bi.

When increasing the Sb/Ga ratio to above stoichiometric (phase II), clear Pendellösung fringes emerge, indicating excellent crystal quality with homogeneous composition. Simultaneously, a maximum of 8.2 %-Bi is reached at Sb/Ga of ~1.03. However, as reported in literature [19, 27], further increase of the Sb/Ga ratio leads to Bi content reduction and, as observed in the last two diffractograms in figure 1(f), crystalline degradation. Our earlier analysis of figure 1(e) containing Bi droplets is consistent the reduced Bi content. In fact, the reduced Bi content in phases I and III is consistent with Tait et al.'s III-V-Bi growth model [40], in which droplet nucleation results in a loss-rate of Bi from the surfactant and thereby reduces the amount of Bi in the crystal termination layer, which governs Bi incorporation. As a whole, the analysis of figure 1(c-f) demonstrates the critical role of Sb/Ga in GaSbBi growth and points to a definite growth window (phase II) with optimal structural quality.

To further map the growth parameter space, all samples in sets A, B and C were characterized with HR-XRD and SEM. The results are compiled in figure 2 and 3 in terms of Bi content, crystalline quality, and surface condition. Due to the low amount of HR-XRD data points, the underlying contour plots in figure 2 and 3 have been interpolated with thin-plate spline (with and without smoothing) and extrapolated to a square grid. The phase II-III interface for sets A and B is plotted as a linear fit to the interface Sb/Ga values of individual samples for simplicity. Also note that the surface morphology in phases I and II are consistent with figure 1(c-d) for all samples, while the phase III droplet character is modified by $T_g$, Sb/Ga and the chosen Bi/Ga BEPR.

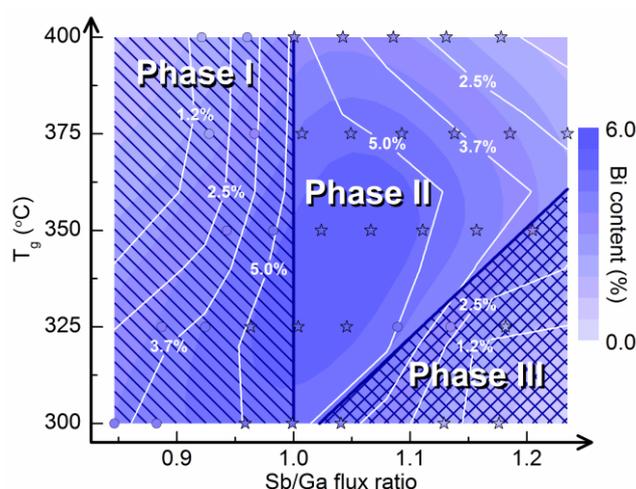

**Figure 2.** Bi content & surface phase vs $T_g$ and Sb/Ga ratio for sample set A. The scatter points represent HR-XRD measurements, with good (stars) and bad (circles) crystalline quality. The shaded regions depict the surface phases. The white contour lines represent unsmoothed data while the colored contours are smoothed.

Figure 2 comprises results for set A with the lowest Bi flux. Phase I has lower Bi content and degraded crystalline quality (circles) confirming our earlier HR-XRD analysis for a range of $T_g$ with lower Bi flux. Correspondingly, moving to phase II increases the Bi content and largely ensures crystal quality (stars). Further increasing Sb/Ga reduces the Bi content although with a temperature dependence. For $T_g$>350 °C phase III is not observed due to the limited Sb/Ga range over the wafer. At these temperatures, Bi surface desorption is higher, which can qualitatively explain the reduced Bi content and shifted phase interface. For $T_g \leq 350$ °C, the reduction in Bi content with decreasing temperature at large Sb/Ga values cannot be explained by a simple kinetic dopant-like incorporation model as used in references [16, 19]. Instead, this could be understood qualitatively through Lewis et al.'s growth model [27], where the Bi incorporation rate is governed by surface stoichiometry and Bi thermal ejection. The latter effect obviously diminishes with decreasing temperature, which should contribute positively to the incorporation, thus not explaining the trend. The former contributor is influenced heavily by the surface kinetics of Ga and Sb. In particular, the Sb desorption rate decreases with $T_g$ resulting into low Ga coverages, which is proportional to Bi incorporation in Lewis' model [27], thus explaining the reduction of Bi incorporation with lower $T_g$. Still, the Bi incorporation wrt. Sb/Ga in figure 2 deviates from Lewis' model, which decreases strongly above the stoichiometric Sb/Ga value. The kinetics of Bi incorporation can be expected to be more complex, albeit controlled by surface stoichiometry. For example, the low vapor pressure of Sb could influence the Bi surfactant layer, which is not considered in GaAsBi based models [25, 27].

The map in figure 2 is reproduced for sets B and C in figure 3. Set B grown with the intermediate Bi flux shows the same trends as set A. The surface phases are situated at similar parameter regimes and Bi content is maximized at intermediate temperatures around the stoichiometric Sb/Ga condition, while phases I and III coincide with lower Bi content and, in general, poor crystalline quality. The most important distinction between sets A and B can be seen in the transition of phase II to a more narrow Sb/Ga range. This growth window closing can be considered as a consequence of reaching the maximum Bi incorporation rate (dependent on Sb/Ga and $T_g$) when increasing the Bi flux. This effect has been reported for single V/III and $T_g$ values for GaSbBi [34] and GaAsBi [24], which here is demonstrated to be critically dependent on Sb/Ga and $T_g$.

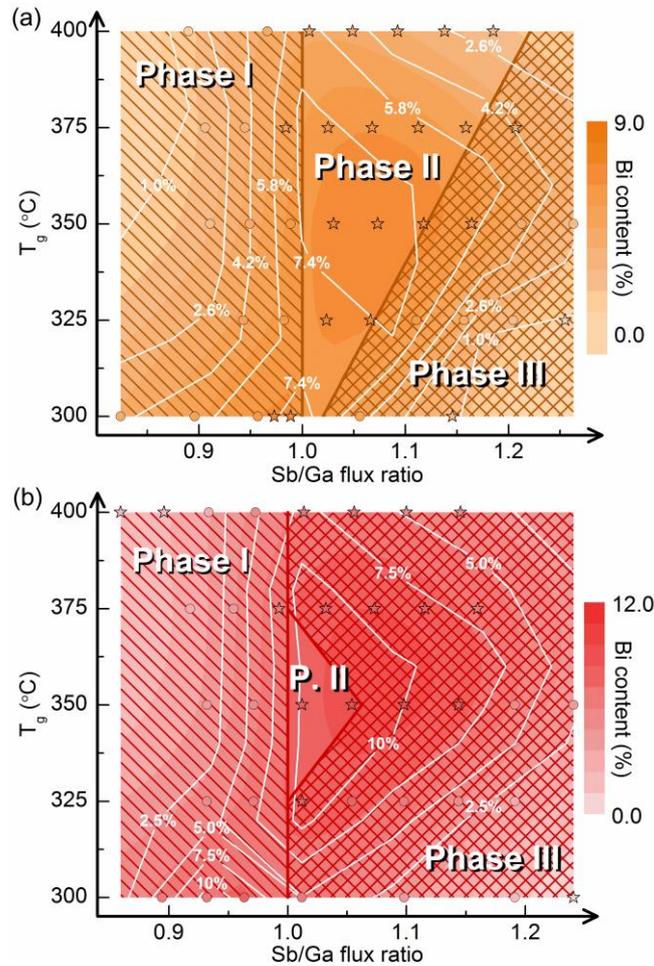

**Figure 3.** Bi content & surface phase vs $T_g$ and Sb/Ga ratio for sample sets B (a) and C (b). The figure components are the same as in figure 2. The phase II-III interface for set B is extrapolated for the highest temperature (400 °C) as it was not observed due to limited Sb/Ga range over the wafer.

Figure 3(b) compiles the data for set C grown with the highest Bi flux, which shows that the growth window has closed for almost every Sb/Ga and $T_g$. In fact, the window is only observed for $T_g$=350 °C, where the window spans Sb/Ga values of ~1–1.06. The now closed window at higher $T_g$ indicates that the Bi desorption and incorporation rates are insufficient to compensate the incident Bi flux. Surprisingly, crystal quality is maintained in phase III for $T_g \geq 350$ °C, which indicates that it can be influenced by thermal effects, such as increased droplet/adatom mobility. We note that some exceptions between crystal quality and surface phase do exist in figure 2 and 3. However, we consider these points as outliers as they are found in areas with low Bi content, which translates into reduced compositional inhomogeneity contrast as well as a narrow range of interpretable HR-XRD features.

The maximum Bi content while maintaining good surface quality was 11.2 % for the 350 °C sample in figure 3(b). This is within 0.2 % of the value that Delorme et al. [19] reported for droplet-free, high crystalline quality GaSbBi films. Similar to Delorme's results, our attempt to increase the Bi content at 350 °C by increasing the Bi/Ga BEPR to 0.18 resulted in a closed phase II window. Near stoichiometric Sb/Ga, this sample showed excellent crystalline quality with 14.5 %-Bi along with Ga-Bi compound droplets of <100 nm size with a density of ~1E8 $cm^{-2}$ (not shown).

Although these results indicate an upper limit of Bi incorporation for droplet-free growth, several methods can be used to prevent droplet formation. Lu et al. [25] originally suggested growth rate to be important in controlling Bi droplet formation in GaAsBi films. Namely, Lu argued that low growth rates would compensate low Bi evaporation rates, thus inhibiting droplet formation. Conversely, Ptak et al. [24] suggested high growth rates to kinetically restrict Bi nucleation. However, Ptak worked in a regime where Bi incorporation was much like a unity-sticking-coefficient dopant, which as observed here is not strictly applicable in general. As with growth rate, we expect that other growth parameters, such as choice in group V species [41] or use of high-index substrates [42, 43], will affect the droplet formation via altered kinetics.

The choice in growth process steps can also inhibit droplet formation. For example, growth interrupts aid desorption of excess Bi from the growth surface. Since the amount of excess Bi is proportional to layer thickness, interrupts have been mainly applied in quantum-well structures [44, 45] where shorter interruption times are required. Additionally, alternate sequencing of fluxes, i.e. migration-enhanced epitaxy (MEE), allows for precise control over the surface stoichiometry and thereby Bi accumulation. So far, few studies exist on MEE-grown GaAsBi [46–48], which have demonstrated high structural and optical quality.

**Acknowledgements**

The work was carried out as a part of EU Horizon 2020 program MIREGAS (grant no. 644192) and Academy of Finland Key project MIRLIGHT (grant no. 305946).

**References**

[1] V. A. Serebryakov, É. V. Boĭko, N. N. Petrishchev, A. V. Yan, "Medical applications of mid-IR lasers. Problems and prospects," *Journal of Optical Technology,* vol. 77, pp. 6-17, 2010.

[2] M. Hermes, R. B. Morrish, L. Huot, L. Meng, S. Junaid, J. Tomko, G. R. Lloyd, W. T. Masselink, P. T. Tidemand-Lichtenberg, C. Pedersen, F. Palombo, N. Stone, "Mid-IR hyperspectral imaging for label-free histopathology and cytology," *Journal of Optics,* vol. 20, p. 023002, 2018.

[3] A. B. Seddon, "Mid-infrared (IR) - A hot topic: The potential for using mid-IR light for non-invasive early detection of skin cancer in vivo," *Physica Status Solidi B,* vol. 250, pp. 1020-1027, 2013.

[4] P. Werle, "Spectroscopic trace gas analysis using semiconductor diode lasers," *Spectrochimica Acta Part A,* vol. 52, pp. 805-822, 1996.

[5] C. S. Goldenstein, R. M. Spearrin, J. B. Jeffries, R. K. Hanson, "Infrared laser-absorption sensing for combustion gases," *Progress in Energy and Combustion Science,* vol. 60, pp. 132-176, 2016.

[6] M. Yin, A. Krier, S. Krier, R. Jones, P. Carrington, "Mid-infrared diode lasers for free-space optical communications," *Proc. SPIE 6399,* p. 63990C, 2006.

[7] R. McClintock, A. Haddadi, M. Razeghi, "Free-space optical communication using mid-infrared or solar-blind ultraviolet sources and detectors," *Proc. SPIE 8268,* p. 826810, 2012.

[8] I. E. Gordon, L. S. Rothman, C. Hill, R. V. Kochanov, Y. Tan, P. F. Bernath, M. Birk, V. Boudon, A. Campargue, K. V. Chance, B. J. Drouin, J.-M. Flaud, R. R. Gamache, J. T. Hodges, D. Jacquemart, V. I. Perevalov, A. Perrin, K. P. Shine, M.-A. H. Smith, J. Tennyson, G. C. Toon, H. Tran, V. G. Tyuterev, A. Barbe, A. G. Császár, V. M. Devi, T. Furtenbacher, J. J. Harrison, J.-M. Hartmann, A. Jolly, T. J. Johnson, T. Karman, I. Kleiner, A. A. Kyuberis, J. Loos, O. M. Lyulin, S. T. Massie, S. N. Mikhailenko, N. Moazzen-Ahmadi, H. S. P. Müller, O. V. Naumenko, A. V. Nikitin, O. L. Polyansky, M. Rey, M. Rotger, S. W. Sharpe, K. Sung, E. Starikova, S. A. Tashkun, J. Vander Auwera, G. Wagner, J. Wilzewski, P. Wcisło, S. Yu, E. J. Zak, "The HITRAN2016 molecular spectroscopic database," *Journal of Quantitative Spectroscopy & Radiative Transfer,* vol. 203, pp. 3-69, 2017.

[9] R. Wang, A. Vasiliev, M. Muneeb, A. Malik, S. Sprengel, G. Boehm, M.-C. Amann, I. Šimonytė, A. Vizbaras, K. Vizbaras, R. Baets, G. Roelkens, "III-V-on-Silicon photonic integrated circuits for spectroscopic sensing in the 2-4 µm wavelength range," *Sensors,* vol. 17, p. 1788, 2017.

[10] R. Soref, "Mid-infrared photonics in silicon and germanium," *Nature Photonics,* vol. 4, pp. 495-497, 2010.

[11] S. D. Sifferman, H. P. Nair, R. Salas, N. T. Sheehan, S. J. Maddox, A. M. Crook, S. R. Bank, "Highly strained mid-infrared type-I diode lasers on GaSb," *IEEE Jounral of Selected Topics in Quantum Electronics,* vol. 21, p. 1502410, 2015.

[12] W. Lei, C. Jagadish, "Lasers and photodetectors for mid-infrared 2-3 µm applications," *Journal of Applied Physics,* vol. 104, p. 091101, 2008.

[13] T. Hosoda, G. Kipshidze, G. Tsvid, L. Shterengas, G. Belenky, "Type-I GaSb-based laser diodes operating in 3.1- to 3.3-µm wavelength range," *IEE Photonics Technology Letters,* vol. 22, pp. 718-720, 2010.

[14] L. Shterengas, G. Belenky, T. Hosoda, G. Kipshidze, S. Suchalkin, "Continous wave operation of diode lasers at 3.36 µm at 12 C," *Applied Physics Letters,* vol. 93, p. 011103, 2008.

[15] G. Belenky, L. Shterengas, G. Kipshidze, T. Hosoda, "Type-I diode lasers for spectral region above 3 µm," *IEEE Journal of Selected Topics in Quantum Electronics,* vol. 17, pp. 1426-1434, 2011.

[16] M. K. Rajpalke, W. M. Linhart, M. Birkett, K. M. Yu, J. Alaria, J. Kopaczek, R. Kudrawiec, T. S. Jones, M. J. Ashwin, T. D. Veal, "High Bi content GaSbBi alloys," *Journal of Applied Physics,* vol. 116, p. 043511, 2014.

[17] L. Yue, X. Chen, Y. Zhang, F. Zhang, L. Wang, J. Shao, S. Wang, "Molecular beam epitaxy growth and optical properties of high bismuth content GaSbBi thin films," *Journal of Alloys and Compounds,* vol. 742, pp. 780-789, 2018.

[18] M. Gladysiewicz, R. Kudrawiec, M. S. Wartak, "Electronic band structure and material gain of III-V-Bi quantum wells grown on GaSb substrate and dedicated for mid-infrared spectral range," *Journal of Applied Physics,* vol. 119, p. 075701, 2016.

[19] O. Delorme, L. Cerutti, E. Tournié, J.-B. Rodriguez, "Molecular beam epitaxy and characterization of high Bi content GaSbBi alloys," *Journal of Crystal Growth,* vol. 477, pp. 144-148, 2017.

[20] D. P. Samajdar, T. D. Das, S. Dhar, "Valence band anticrossing model for GaSbBi and GaPBi using k.p method," *Materials Science in Semiconductor Processing,* vol. 40, pp. 539-542, 2015.

[21] O. Delorme, L. Cerutti, E. Luna, G. Narcy, A. Trampert, E. Tournié, J.-B. Rodriguez, "GaSbBi/GaSb quantum well laser diodes," *Applied Physics Letters,* vol. 110, p. 222106, 2017.

[22] L. Wang, L. Zhang, L. Yue, D. Liang, X. Chen, Y. Li, P. Lu, J. Shao, S. Wang, "Novel Dilute Bismide, Epitaxy, Physical Properties and Device Application," *Crystals,* vol. 7, p. 63, 2017.

[23] S. Tixier, M. Adamcyk, T. Tiedje, S. Francoeur, A. Mascarenhas, P. Wei, F. Schiettekatte, "Molecular beam epitaxy growth of GaAsBi," *Applied Physics Letters,* vol. 82, p. 2245, 2003.

[24] A. J. Ptak, R. France, D. A. Beaton, K. Alberi, K. Simon, A. Mascarenhas, C.-S. Jiang, "Kinetically limited growth of GaAsBi by molecular-beam epitaxy," *Journal of Crystal Growth,* vol. 338, p. 107, 2011.

[25] X. Lu, D. A. Beaton, R. B. Lewis, T. Tiedje, M. B. Whitwick, "Effect of molecular beam epitaxy growth conditions on the Bi content of GaAsBi," *Applied Physics Letters,* vol. 92, p. 192110, 2008.

[26] A. R. Mohmad, F. Bastiman, C. J. Hunter, J. S. Ng, S. J. Sweeney, J. P. R. David, "The effect of Bi composition to the optical quality of GaAsBi," *Applied Physics Letters,* vol. 99, p. 042107, 2011.

[27] R. B. Lewis, M. Masnadi-Shirazi, T. Tiedje, "Growth of high Bi concentration GaAsBi by molecular beam epitaxy," *Applied Physics Letters,* vol. 101, no. 8, p. 082112, 2012.

[28] G. Vardar, S. W. Paleg, M. V. Warren, M. Kang, S. Jeon, R. S. Goldman, "Mechanisms of droplet formation and Bi incorporation during molecular beam epitaxy of GaAsBi," *Applied Physics Letters,* vol. 102, p. 042106, 2013.

[29] J. Puustinen, J. Hilska, M. Guina, "Analysis of GaAsBi growth regimes in high resolution with respect to As/Ga ratio using stationary MBE growth," *in press, Journal of Crystal Growth.*

[30] F. Tsui, L. He, "Techniques for combinatorial molecular beam epitaxy," *Review of Scientific Instruments,* vol. 76, p. 062206, 2005.

[31] A. Duzik, J. M. Millunchick, "Surface morphology and Bi incorporation in GaSbBi(As)/GaSb films," *Journal of Crystal Growth,* vol. 390, pp. 5-11, 2014.

[32] E. Sterzer, N. Knaub, P. Ludewig, R. Straubinger, A. Beyer, K. Volz, "Investigation of the microstructure of metallic droplets on Ga(AsBi)/GaAs," *Journal of Crystal Growth,* vol. 408, pp. 71-77, 2014.

[33] M. Masnadi-Shirazi, R. B. Lewis, V. Bahrami-Yekta, T. Tiedje, M. Chicoine, P. Servati, "Bandgap and optical absorption edge of GaAsBi alloys with $0 < x < 17.8\ \%$," *Journal of Applied Physics,* vol. 116, p. 223506, 2014.

[34] M. K. Rajpalke, W. M. Linhart, K. M. Yu, T. S. Jones, M. J. Ashwin, T. D. Veal, "Bi flux-dependent MBE growth of GaSbBi alloys," *Journal of Crystal Growth,* vol. 425, pp. 241-244, 2015.

[35] B. Z. Nosho, B. R. Bennett, E. H. Aifer, M. Goldenberg, "Surface morphology of homoepitaxial GaSb films grown on flat and vicinal substrates," *Journal of Crystal Growth,* vol. 236, pp. 155-164, 2002.

[36] G. Apostolopoulos, N. Boukos, J. Herfort, A. Travlos, K. H. Ploog, "Surface morphology of low temperature grown GaAs on singular, vicinal substrates," *Materials Sceince, Engineering,* vol. B88, pp. 205-208, 2002.

[37] M. K. Rajpalke, W. M. Linhart, K. M. Yu, M. Birkett, J. Alaria, J. J. Bomphrey, S. Sallis, L. F. J. Piper, T. S. Jones, M. J. Ashwin, T. D. Veal, "Bi-induced band gap reduction in epitaxial InSbBi alloys," *Applied Physics Letters,* vol. 105, p. 212101, 2014.

[38] J. A. Steele, R. A. Lewis, "Laser-induced oxidation kinetics of bismuth surface microdroplets on GaAsBi studied in situ by Raman microprobe analysis," *Optics Express,* vol. 22, no. 26, pp. 32261-32275, 2014.

[39] C. R. Tait, L. Yan, J. M. Millunchick, "Droplet induced compositional inhomogeneities in GaAsBi," *Applied Physics Letters,* vol. 111, p. 042105, 2017.

[40] C. R. Tait, J. M. Millunchick, "Kinetics of droplet formation and Bi incorporation in GaSbBi alloys," *Journal of Applied Physics,* vol. 119, p. 215302, 2016.

[41] R. D. Richards, F. Bastiman, C. J. Hunter, D. F. Mendes, A. R. Mohmad, J. S. Roberts, J. P. R. David, "Molecular beam epitaxy growth of GaAsBi using As2 and As4," *Journal of Crystal Growth,* vol. 390, p. 120, 2014.

[42] M. Henini, J. Ibáñez, M. Schmidbauer, M. Shafi, S. V. Novikov, L. Turyanska, S. I. Molina, D. L. Sales, M. F. Chisholm, J. Misiewicz, "Molecular beam epitaxy of GaBiAs on (311)B GaAs substrates," *Applied Physics Letters,* vol. 91, p. 251909, 2007.

[43] J. Li, K. Collar, W. Jiao, W. Kong, T. F. Kuech, S. E. Babcock, A. Brown, "Impact of vicinal GaAs(001) substrates on Bi incorporation and photoluminescence in molecular beam epitaxy-grown GaAsBi," *Applied Physics Letters,* vol. 108, p. 232102, 2016.

[44] R. D. Richards, F. Bastiman, D. Walker, R. Beanland, J. P. R. David, "Growth and structural characterization of GaAsBi/GaAs multiple quantum wells," *Semiconductor Science and Technology,* vol. 30, p. 094013, 2015.

[45] P. K. Patil, E. Luna, T. Matsuda, K. Yamada, K. Kamiya, F. Ishikawa, S. Shimomura, "GaAsBi/GaAs multi-quantum well LED grown by molecular beam epitaxy using a two-substrate-temperature technique," *Nanotechnology,* vol. 28, p. 105702, 2017.

[46] R. Butkutė, V. Pačebutas, A. Krotkus, N. Knaub, K. Volz, "Migration-enhanced epitaxy of thin GaAsBi layers," *Lithuanian Journal of Physics,* Vols. 125-129, p. 54, 2014.

[47] R. Butkutė, K. Stašys, V. Pačebutas, B. Čechavičius, R. Kondrotas, A. Geižutis, "Bismuth quantum dots and strong infrared photoluminescence in migration-enhanced epitaxy grown GaAsBi-based structures," *Optical and Quantum Electronics,* vol. 47, pp. 873-882, 2015.

[48] V. Pačebutas, R. Butkutė, B. Čechavičius, S. Stanionytė, E. Pozingytė, M. Skapas, A. Selskis, A. Geižutis, A. Krotkus, "Bismides: 2D structures and quantum dots," *Journal of Physics D: Applied Physics,* vol. 50, p. 364002,

2017.